\documentclass[aps,prd,twocolumn,showpacs]{revtex4}
\usepackage{amsmath}

\newcommand{\mean}[1]{\langle #1 \rangle} 

\begin{document}

\preprint{Saclay-T03/017}

\title{Multiparticle correlations from momentum conservation}

\author{Nicolas Borghini}
\email{borghini@spht.saclay.cea.fr}
\affiliation{Service de Physique Th{\'e}orique, CEA-Saclay, 
F-91191 Gif-sur-Yvette cedex, France}
\date{\today}

\begin{abstract}
Using a generating-function formalism, we compute the contribution of momentum 
conservation to multiparticle correlations between the emitted particles in 
high-energy collisions. 
In particular, we derive a compact expression of the genuine $M$-particle 
correlation, for arbitrary $M$. 
\end{abstract}


\maketitle

The main purpose of this paper is to deal with a rather general problem: 
consider $N$ random variables (in a $D$-dimensional space) ${\bf p}_1$, \ldots, 
${\bf p}_N$, which are independent except for a constraint 
${\bf p}_1 + \ldots + {\bf p}_N={\bf 0}$. 
What is the multiple correlation between $M$ variables among the ${\bf p}_j$ 
(with $M<N$) induced by this constraint, and in particular, what is the 
{\em cumulant} of the $M$-variable correlation?

One may for example think of the correlation between $M$ monomers of a 
finite-size-$N$ ring polymer.
Another instance of the problem under study, from which we borrow the 
terminology in this paper, is the constraint arising from global momentum 
conservation in high-energy collisions: in the center-of-mass frame, the sum 
of the momenta of the $N$ emitted particles vanishes, and this induces 
correlations between $M$ arbitrary particles. 

An accurate knowledge of these unavoidable multiparticle correlations due to 
global momentum conservation is important. 
They represent an ever-present background effect to other sources of 
correlations between the outgoing particles, which are a primary target of 
investigation, as they provide information on the physics involved in the 
collisions~\cite{DeWolf:1995pc}: 
either short-range correlations, as quantum (Bose--Einstein or Fermi--Dirac) 
effects~\cite{Boal:yh,HBT-HIC2p,HBT-HIC3p}, correlations between decay 
products, (mini)jets, etc.;
or inter-particle correlations arising from a collective motion (``flow'') in 
the context of heavy-ion collisions~\cite{JY:QM97}.
For instance, it has been shown that the analysis of collective flow can be 
biased by the two-particle correlations due to global momentum 
conservation~\cite{Borghini:2000cm}

A common property of short-range correlations is their scaling. 
If $N$ denotes the total number of emitted particles, the connected part of 
the $M$-particle correlation scales as $1/N^{M-1}$. 
This scaling follows from simple combinatorics: $1/N^{M-1}$ is the probability 
that $M$ arbitrary particles have been together in a small phase-space region. 

Global momentum conservation is certainly not a short-range correlation, since 
it affects {\em all} particles in the collision.
One may therefore wonder how the resulting correlation scales with $N$ for a
given $M$. 
The two-particle case has already been 
calculated~\cite{Borghini:2000cm,Danielewicz:1988in}. 
In the following, we extend the calculation to $M$-particle correlations for 
arbitrary $M$, using a generating function of the correlations and a 
saddle-point approximation. 
In particular, we shall derive a compact expression for the correlation at 
any order $M$, to leading order in $1/N$, and show that it follows the same 
behavior as short-range correlations. 

We first introduce, in Sec.~\ref{s:formalism}, a generating function of the 
multiparticle correlations. 
This generating function then allows us to derive the expression of the genuine 
correlation (cumulant) between an arbitrary number of particles to leading 
order in $1/N$ (Sec.~\ref{s:order}), and in particular the scaling of the 
cumulants. 
As an illustration of this result, we then compute explicitly the two- and 
three-particle correlations (Sec.~\ref{s:c2,c3}).
The results are commented in Sec.~\ref{s:conclusion}.

\section{Multiparticle cumulants}
\label{s:formalism}

Consider a collision in which a total of $N$ particles are emitted (throughout
this paper, $N$ is assumed to be large).
Let ${\bf p}_1$, \ldots, ${\bf p}_N$, be their $D$-dimensional momenta, where 
the space dimension $D$ is kept arbitrary for sake of generality. 
For a given $M<N$, we denote by $f({\bf p}_{j_1}, \ldots, {\bf p}_{j_M})$ the 
$M$-particle probability distribution of ${\bf p}_{j_1}$, \ldots, 
${\bf p}_{j_M}$. 

To derive multiparticle correlations of arbitrary order in a systematic way, 
we introduce a generating function of the distributions, which we define as
\begin{eqnarray}
\label{genfunc}
G(x_1,\ldots,x_N) &\equiv& 1 + x_1 f({\bf p}_1) + x_2 f({\bf p}_2) + \cdots \cr
 & & \ \ + x_1 x_2 f({\bf p}_1,{\bf p}_2) + \cdots,
\end{eqnarray}
and so on for every order $M$. 
Thus, the coefficient of the term $x_{j_1} x_{j_2} \ldots x_{j_M}$ is the 
$M$-particle probability distribution 
$f({\bf p}_{j_1}, {\bf p}_{j_2}, \ldots, {\bf p}_{j_M})$. 

We shall mostly be interested in the {\em connected part} of the $M$-particle 
distribution, that is, the term which cannot be expressed as a product of 
correlations between less than $M$ particles. 
For instance, the two-particle distribution can be decomposed into two parts: 
\[
f({\bf p}_1,{\bf p}_2) = f({\bf p}_1)\,f({\bf p}_2) + f_c({\bf p}_1,{\bf p}_2),
\]
where the first term in the right-hand side (rhs) is the mere product of the 
one-particle probability distributions, while the second one, the ``connected 
part,'' reflects correlations between ${\bf p}_1$ and ${\bf p}_2$. 
In particular, $f_c({\bf p}_1,{\bf p}_2)$ vanishes if ${\bf p}_1$ and 
${\bf p}_2$ are uncorrelated.
This genuine correlation, which is also called the {\it cumulant} of the 
distribution in probability theory, is obtained by taking the logarithm of the 
generating function of distributions~\cite{vanKampen}. 
The cumulant $f_c({\bf p}_{j_1}, \ldots, {\bf p}_{j_M})$ is the coefficient of 
$x_{j_1}\ldots x_{j_M}$ in the expansion of $\ln G(x_1, \ldots, x_N)$:
\begin{eqnarray}
\label{genfunccum}
\ln G(x_1,\ldots,x_N) &\!\!\equiv\!\!& 
x_1 f_c({\bf p}_1) + x_2 f_c({\bf p}_2) + \cdots \cr
 & & \qquad\qquad +\, x_1 x_2 f_c({\bf p}_1,{\bf p}_2) + \cdots.\quad
\end{eqnarray}

If the system splits into independent subsystems or, more generally, if  
correlations in the system are short-range, the probability distributions of 
each subsystem add up: 
$f(\{{\bf p}_j\}) = \sum_A (N_A/N) f_A(\{{\bf p}_j\})$, where the sum runs 
over subsystems with sizes $N_A$. 
It follows that $G$ can be factorized into the product of functions for each 
subsystem: $G(\{x_j\}) = \prod_A g_A(\{N_Ax_j/N\})$.
In the limit where each individual particle is a subsystem, 
$G(\{x_j\}) = [g(\{x_j/N\})]^N$. 
Therefore, $\ln G$, and the cumulants, are the sums of the corresponding 
quantities for the subsystems. 
As a result, inspecting the term in $x_{j_1}\ldots x_{j_M}$ shows that, while 
the $M$-particle probability distribution 
$f({\bf p}_{j_1}, \ldots, {\bf p}_{j_M})$ is independent of $N$, the cumulant 
$f_c({\bf p}_{j_1}, \ldots, {\bf p}_{j_M})$ scales like $1/N^{M-1}$. 
In Sec.~\ref{s:order}, we shall show that the cumulants from the correlations 
due to global momentum conservation follow the same scaling to leading order 
in $1/N$. 
More precisely, we shall demonstrate that the generating function of 
distributions reads 
\[
G(x_1,\ldots,x_N) \propto e^{Ng(\{x_j/N\})} 
\left( 1 + \sum \frac{\{x_j\}^l}{N^q} \right),
\]
where the sum runs over terms with $q \geq l$.

\section{Cumulants in the large-$N$ limit}
\label{s:order}

In this Section, we derive a compact expression of the cumulants of 
multiparticle correlations due to global momentum conservation to leading order 
in $1/N$. 
In particular, we show that the $M$-particle cumulants scale like $1/N^{M-1}$. 

Let us assume for simplicity that the only source of correlation between the 
particles is global momentum conservation, ${\bf p}_1+\cdots+{\bf p}_N={\bf 0}$. 
Under this assumption, the $M$-particle distribution of ${\bf p}_1$, \ldots,
${\bf p}_M$, with $M<N$, is defined as
\begin{widetext}
\begin{equation}
\label{def_kdis}
f({\bf p}_1, \ldots, {\bf p}_M) \equiv 
\frac{\displaystyle \left(\prod_{j=1}^M F({\bf p}_j) \right) 
\int \!\delta^D({\bf p}_1+\cdots+{\bf p}_N) 
\prod_{j=M+1}^N \left[ F({\bf p}_j)\,d^D{\bf p}_j \right] / {\cal N}_D^{N-M}}{
\displaystyle \int \!\delta^D({\bf p}_1+\cdots+{\bf p}_N) 
\prod_{j=1}^N \left[ F({\bf p}_j)\,d^D{\bf p}_j \right] / {\cal N}_D^N},
\end{equation}
where $F({\bf p})$ is the one-particle momentum distribution unrenormalized 
for the momentum conservation constraint, and 
${\cal N}_D \equiv \int \! F({\bf p})\,d^D{\bf p}$ is a normalization constant.
In the following, we denote by $\mean{\ldots}$ an $F({\bf p})$-weighted 
average, that is,
$\mean{g({\bf p})}\equiv\int\! g({\bf p}) F({\bf p}) \,d^D{\bf p}/{\cal N}_D$,
for any function of momentum $g({\bf p})$. 

The denominator in Eq.~(\ref{def_kdis}) is a constant, which we denote by 
$1/{\cal C}_D$, independent of $M$.
The actual value will not influence the forthcoming discussion. 

Consider next the numerator of Eq.~(\ref{def_kdis}). 
Introducing a Fourier representation of the Dirac distribution, it reads
\begin{equation}
\label{denom}
\int \!\frac{d^D{\bf k}}{(2\pi)^D}
\left( \prod_{j=1}^M F({\bf p}_j) e^{i{\bf k}\cdot{\bf p}_j} \right)
\mean{e^{i{\bf k}\cdot{\bf p}}}^{N-M} = \int\!\frac{d^D{\bf k}}{(2\pi)^D} \, 
\mean{e^{i{\bf k}\cdot{\bf p}}}^N
\left( \prod_{j=1}^M F({\bf p}_j) 
\frac{e^{i{\bf k}\cdot{\bf p}_j}}{\mean{e^{i{\bf k}\cdot{\bf p}}}} \right). 
\end{equation}

Inserting in Eq.~(\ref{genfunc}) the expression of the $M$-particle 
distribution, Eq.~(\ref{def_kdis}), with the numerator replaced by the 
rhs of Eq.~(\ref{denom}), we obtain 
\begin{eqnarray}
\label{genfunc2}
G(x_1,\ldots,x_N) & = & {\cal C}_D
\int\!\frac{d^D{\bf k}}{(2\pi)^D} \, \mean{e^{i{\bf k}\cdot{\bf p}}}^N
\prod_{j=1}^N \left( 1+ x_j F({\bf p}_j) 
\frac{e^{i{\bf k}\cdot{\bf p}_j}}{\mean{e^{i{\bf k}\cdot{\bf p}}}} \right) \cr
 & \simeq & {\cal C}_D
\int\!\frac{d^D{\bf k}}{(2\pi)^D} \, \mean{e^{i{\bf k}\cdot{\bf p}}}^N
\exp\!\left( \sum_{j=1}^N x_j F({\bf p}_j) 
\frac{e^{i{\bf k}\cdot{\bf p}_j}}{\mean{e^{i{\bf k}\cdot{\bf p}}}} \right). 
\end{eqnarray}
\end{widetext}
In passing from the first line to the second one, we have used the fact that 
we shall only consider the coefficient of $x_{j_1}\ldots x_{j_M}$ where the 
$M$ indices $j_1$, \ldots, $j_M$ are all different [see 
Eq.~(\ref{genfunccum})].
One easily checks that the only difference between the two forms of the 
generating function in Eq.~(\ref{genfunc2}) comes from terms in which at least 
one $x_j$ is raised to some power $m\geq 2$ (corresponding to the 
``autocorrelation'' of particle $j$ with itself). 
Therefore, as far as we are concerned, Eq.~(\ref{genfunc2}) really is an 
{\em identity}, not an approximation.

Please note that in Eq.~(\ref{genfunc2}), the variable $x_j$ is always 
multiplied by a factor $F({\bf p}_j)$. 
We may thus rescale $x_j$ by this factor, and drop it in the following to 
simplify expressions. 
This is quite satisfactory, since it means that the measurable multiparticle 
correlations $f({\bf p}_{j_1}, \ldots, {\bf p}_{j_M})$ or 
$f_c({\bf p}_{j_1}, \ldots, {\bf p}_{j_M})$ will not depend on the 
non-measurable distribution $F({\bf p})$---this justifies our previously 
calling it ``unrenormalized''. 

To evaluate the integral in Eq.~(\ref{genfunc2}), we rely on the fact that $N$ 
is large, and use a saddle-point approximation. 
To simplify the discussion, we introduce the notation
\begin{equation}
\label{F(k)}
{\cal F}({\bf k}) \equiv \ln \mean{e^{i{\bf k}\cdot{\bf p}}} + 
\sum_{j=1}^N \frac{x_j}{N}
\frac{e^{i{\bf k}\cdot{\bf p}_j}}{\mean{e^{i{\bf k}\cdot{\bf p}}}}. 
\end{equation}
Thus, the integrand in Eq.~(\ref{genfunc2}) is $e^{N{\cal F}({\bf k})}$. 
We call ${\bf k}_0$ the position of the saddle point.
Please note that ${\cal F}$ and its successive derivatives, which we shall 
denote by ${\cal F}'$, ${\cal F}''$, ${\cal F}^{(3)}$, \ldots, depend on $x$ 
only through $x/N$, where $x$ stands for any of the $x_j$. 
Therefore, ${\bf k}_0$, which is of course the solution of 
${\cal F}'({\bf k})=0$, also depends on $x/N$ only. 
One should pay attention to the fact that the saddle point ${\bf k}_0$ is not 
merely the origin ${\bf k}={\bf 0}$.

We shall now demonstrate that to leading order in $1/N$, all multiparticle 
cumulants are determined by the saddle-point value $N{\cal F}({\bf k}_0)$. 
More precisely, we show that the $M$-particle cumulant is of order $1/N^{M-1}$, 
and that corrections to the saddle-point calculation only yield subleading 
terms, suppressed by (positive) powers of $1/N$. 

Using a Taylor expansion of ${\cal F}({\bf k})$ around the saddle point, the 
generating function (\ref{genfunc2}) reads
\begin{widetext}
\begin{eqnarray}
\label{genfunc3}
G(x_1,\ldots,x_N) & = & {\cal C}_D\,e^{N{\cal F}({\bf k}_0)}
\int\!\frac{d^D{\bf k}}{(2\pi)^D} \, 
e^{N {\cal F}''({\bf k}_0) ({\bf k}-{\bf k}_0)^2/2}\,
\exp\!\left[ N\sum_{m=3}^{+\infty} \frac{{\cal F}^{(m)}({\bf k}_0)}{m!}
({\bf k}-{\bf k}_0)^m \right]\cr
 & = & {\cal C}_D\,e^{N{\cal F}({\bf k}_0)}
\int\!\frac{d^D{\bf k}}{(2\pi)^D} \, 
e^{N {\cal F}''({\bf k}_0) ({\bf k}-{\bf k}_0)^2/2} 
\left( 1+ \sum_{n=1}^{+\infty} \frac{1}{n!} \left[N \sum_{m=3}^{+\infty} 
\frac{{\cal F}^{(m)}({\bf k}_0)}{m!} ({\bf k}-{\bf k}_0)^m \right]^n \right).
\end{eqnarray}
\end{widetext}

As recalled in Sec.\ \ref{s:formalism}, the cumulants are given by the 
logarithm of the generating function. 
Now, the logarithm of Eq.~(\ref{genfunc3}) will split into three parts. 
First, there is $\ln {\cal C}_D$, which does not depend on $x$, and thus does
not influence the values of the multiparticle cumulants. 

Next, there is $N{\cal F}({\bf k}_0)$. 
As noted above, both ${\cal F}$ and ${\bf k}_0$ only involve powers of $x/N$. 
Thus, the coefficient of $x_{j_1}\ldots x_{j_M}$ in ${\cal F}({\bf k}_0)$ 
contains a factor $1/N^M$. 
Multiplying by the overall factor of $N$, we find that the contribution of 
$N{\cal F}({\bf k}_0)$ to the cumulant $f_c(x_{j_1}\ldots x_{j_M})$ scales as
$1/N^{M-1}$. 

Finally, the cumulants involve the contribution of the logarithm of the 
integral in Eq.~(\ref{genfunc3}).
We shall use the fact that the integrand is the sum of a Gaussian function 
and of combinations of its moments. 
After integration, this yields a common factor, 
$1/[2\pi N {\cal F}''({\bf k}_0)]^{D/2}$, which multiplies a sum $\Sigma$. 
Let us show that $\Sigma$ only involves terms in $x^l/N^q$ with $q \geq l$. 
Noting that the values which contribute to the Gaussian integral are of order 
$({\bf k}-{\bf k}_0)^2 \approx 1/N$, one sees that each term 
$N{\cal F}^{(m)}({\bf k}_0) ({\bf k}-{\bf k}_0)^m$ is of order $N^{1-m/2}$, 
multiplied by a function of $x/N$. 
Since the sum runs over $m\geq 3$, $N^{1-m/2} \leq N^{-1/2}$: therefore, 
a power $x^l$ goes with at most a factor $1/N^{l+1/2}$. 
Even after raising the sum to the power $n$, and taking the logarithm, this 
will always remain of the form $x^l/N^q$ with $q > l$.
Finally, the logarithm of the overall factor 
$1/[2\pi N {\cal F}''({\bf k}_0)]^{D/2}$ involves $x$ only through the form 
of powers of $x/N$ in $\ln {\cal F}''({\bf k}_0)$. 
All in all, the contribution to the $M$-particle cumulant of the integral is 
therefore at most of order $1/N^M$, subleading with respect to the contribution 
of $N{\cal F}({\bf k}_0)$.

Therefore, we have shown that the leading contribution to the cumulants of 
multiparticle correlations due to momentum conservation is the saddle-point 
value: $\ln G(x_1,\ldots,x_N) = N{\cal F}({\bf k}_0)$.
This means first that the mere knowledge of ${\bf k}_0$ gives access to all 
cumulants at once, at least to leading order. 
One easily checks that ${\bf k}_0$ is pure imaginary, i.e., $i{\bf k}_0$ has 
real-valued components. 
{}From Eq.~(\ref{F(k)}), it follows that ${\cal F}({\bf k}_0)$ is real-valued 
as well, and so are the cumulants, as should be. 
We shall illustrate these points by computing explicitly the two- and 
three-particle cumulants in the following Section.
Our result also means that genuine $M$-particle correlations arising from 
momentum conservation, which is a long-range effect, scale in the same way as 
correlations from short-range sources, namely as ${\cal O}(1/N^{M-1})$. 
This was certainly not obvious {\em a priori}.

\section{Two- and three-particle cumulants}
\label{s:c2,c3}

As an example of the order of magnitude derived in the previous Section, 
let us compute the two- and three-particle cumulants. 
According to the discussion in Sec.~\ref{s:formalism}, they are given by the 
terms in $x^2$ and $x^3$ in the generating function of cumulants 
$\ln G(x_1, \ldots, x_N)$, that is, following Sec.~\ref{s:order}, the 
corresponding terms in $N{\cal F}({\bf k}_0)$. 
We shall for simplicity perform calculations assuming that $F({\bf p})$ only 
depends on the modulus $|{\bf p}|$, not on the azimuthal angle of ${\bf p}$. 
As a consequence, the average momentum $\mean{\bf p}$ vanishes.
Departures from this assumption are discussed at the end of this Section.

Now a straightforward calculation shows that the saddle point ${\bf k}_0$ is 
given by 
\begin{equation}
\label{F'(k0)}
\left( \sum_{j=1}^N \frac{x_j}{N} 
\frac{e^{i{\bf k}_0\cdot{\bf p}_j}}{\mean{e^{i{\bf k}_0\cdot{\bf p}}}} - 1 
\right) \mean{{\bf p}e^{i{\bf k}_0\cdot{\bf p}}} = 
\sum_{j=1}^N \frac{x_j}{N} {\bf p}_j \,e^{i{\bf k}_0\cdot{\bf p}_j}.
\end{equation}
As mentioned above, ${\bf k}_0$ is a function of $x/N$. 
Equation (\ref{F'(k0)}) can be solved order by order in $x/N$, using the fact 
that one term (in the left-hand side) is of order $0$ in $x/N$ while the other 
two are linear in $x/N$.
Inspecting Eq.~(\ref{F(k)}), and using the fact that the first term in the rhs 
is even in ${\bf k}$, we see that the calculation of ${\cal F}({\bf k}_0)$ to 
order $x^3$ requires our knowing ${\bf k}_0$ to order $x^2$ (while the 
calculation to order $x^2$ only requires ${\bf k}_0$ to order $x$). 

We can then compute ${\bf k}_0$, expanding Eq.~(\ref{F'(k0)}) in powers of 
${\bf k}_0$, which gives 
\begin{equation}
\label{Eq.k0}
i{\bf k}_0 = 
-\left[ 1_D-\left(X_01_D - \frac{D}{\mean{{\bf p}^2}} X_2 \right) \right]^{-1} 
\frac{D}{\mean{{\bf p}^2}} {\bf X_1},
\end{equation}
where $1_D$ denotes the unit $D\times D$ matrix and we have introduced 
\[
X_0 \equiv \sum_{j=1}^N \frac{x_j}{N}, \quad
{\bf X_1} \equiv \sum_{j=1}^N \frac{x_j}{N} {\bf p}_j, \quad
X_2 \equiv \sum_{j=1}^N \frac{x_j}{N} {\bf p}_j \otimes {\bf p}_j.
\]
Note that $i{\bf k}_0$ is real-valued, as expected. 
Reporting its value in Eq.~(\ref{F(k)}), we obtain 
\begin{eqnarray*}
{\cal F}({\bf k}_0) &\!=\!& X_0 - \frac{D}{2\mean{{\bf p}^2}} ({\bf X_1})^2 \cr
 & & \quad\,\ - \frac{D}{2\mean{{\bf p}^2}} {\bf X_1}\cdot 
\left( X_0 1_D - \frac{D}{\mean{{\bf p}^2}} X_2 \right) \cdot {\bf X_1}.
\end{eqnarray*}
Multiplying this result by $N$, we finally obtain, to leading order in $1/N$: 
\begin{widetext}
\begin{eqnarray}
\ln G(x_1, \ldots, x_N) &=& \sum_{j=1}^N x_j - 
\frac{D}{2N \langle {\bf p}^2 \rangle}
\sum_{j,k} x_j x_k ({\bf p}_j \cdot {\bf p}_k) \cr 
 & & \qquad\ \,\ -\ \frac{D}{2N^2\langle {\bf p}^2 \rangle} 
\sum_{j,k,l} x_j x_k x_l \left[ {\bf p}_j \cdot {\bf p}_l  -  
\frac{D}{\langle {\bf p}^2 \rangle} 
({\bf p}_j \cdot {\bf p}_k)({\bf p}_k \cdot {\bf p}_l) \right] + {\cal O}(x^4).
\end{eqnarray}
Hence the coefficients of $x_1 x_2$ and $x_1 x_2 x_3$:
\begin{eqnarray}
\label{c2}
f_c({\bf p}_1, {\bf p}_2) &=& \displaystyle 
-\frac{D\,{\bf p}_1\cdot {\bf p}_2}{N \mean{{\bf p}^2}}, \\
f_c({\bf p}_1, {\bf p}_2, {\bf p}_3) & = & \displaystyle
-\frac{D}{N^2\langle {\bf p}^2 \rangle} 
({\bf p}_1 \cdot {\bf p}_2 + {\bf p}_1 \cdot {\bf p}_3 + {\bf p}_2 \cdot {\bf p}_3) \cr
 & & + \displaystyle \frac{D^2}{N^2\langle {\bf p}^2 \rangle^2} \left[
({\bf p}_1 \cdot {\bf p}_2)({\bf p}_1 \cdot {\bf p}_3) + 
({\bf p}_1 \cdot {\bf p}_2)({\bf p}_2 \cdot {\bf p}_3) + 
({\bf p}_1 \cdot {\bf p}_3)({\bf p}_2 \cdot {\bf p}_3) \right]. 
\label{c3}
\end{eqnarray}
\end{widetext}
We recover, in the case $D=2$, the expression of the two-particle correlation 
due to {\em transverse} momentum conservation already derived in 
Refs~\cite{Borghini:2000cm,Danielewicz:1988in}: the correlation is back-to-back,
and stronger between high-momenta particles.

Finally, let us comment on the assumption that $F({\bf p})$ only depends on 
the modulus of ${\bf p}$, not on its azimuthal angle. 

Our expansion of Eq.~(\ref{F'(k0)}) relies on both $\mean{\bf p}={\bf 0}$ and 
on the identity $\mean{({\bf k}\cdot{\bf p})^2} = {\bf k}^2\mean{{\bf p}^2}/D$, 
which are no longer valid if $F$ is non-isotropic. 
In the general case, ${\bf k}^2 \mean{{\bf p}^2}/D$ will be replaced by another 
quadratic form in ${\bf k}$, namely 
${\bf k} \cdot \mean{{\bf p'} \otimes {\bf p'}} \cdot{\bf k}$, where 
${\bf p'} \equiv {\bf p} -\mean{\bf p}$. 
To perform explicit calculations, it becomes necessary to introduce the 
principal-axes frame, in which ${\bf p'} \otimes {\bf p'}$ is diagonal. 
For instance, the two-particle cumulant, Eq.~(\ref{c2}), will read 
\begin{equation}
\label{c2bis}
f_c({\bf p}_1, {\bf p}_2) = 
- \sum_{i=1}^D \frac{({\bf p}_1)_i ({\bf p}_2)_i}{N \mean{({\bf p})_i^2}},
\end{equation}
where the sum runs over the coordinates along the principal axes.
The distinction between the various directions is for example relevant in a 
high-energy collision: while the {\em transverse} momentum distribution can 
be isotropic (unless there is some anisotropy, as, e.g., a correlation to the 
impact parameter direction), the isotropy will not extend to the beam direction. 
In that case, one may want to use Eq.~(\ref{c2}), with $D=2$, when studying 
two-particle correlations in the transverse plane, but turn to Eq.~(\ref{c2bis})
if interested in $3$-dimensional correlations.

\section{Discussion}
\label{s:conclusion}

We have shown how it is possible to calculate the multiparticle correlations 
arising from momentum conservation between any number of particles using a 
generating function formalism: 
\begin{subequations}
\begin{equation}
\ln G(x_1, \ldots, x_N) = N{\cal F}({\bf k}_0),
\end{equation}
where ${\cal F}({\bf k})$ is given by Eq.~(\ref{F(k)}) and ${\bf k}_0$ by 
\begin{equation}
{\cal F}'({\bf k}_0) = 0.
\end{equation}
\end{subequations}
In particular, we have seen that the $M$-particle cumulant scales as 
$1/N^{M-1}$, where $N$ is the total number of emitted particles.
That means that the correlation scales in the same way as correlations arising 
{}from short-range final interactions or from resonance decays, although the 
underlying reason is less obvious. 
In the latter cases, the scaling is a simple consequence of combinatorics: the 
$M$ particles are required either to be altogether in a small phase-space 
region, or to originate from a single decay, hence the factor $1/N^{M-1}$. 
However, in the case of global momentum conservation, the same arguments do 
not apply {\em a priori}.

The scaling of the genuine $M$-particle correlation due to momentum 
conservation has several consequences. 
First, a good feature: in the context of collective flow in heavy-ion 
collisions~\cite{JY:QM97}, a new method of analysis has been 
proposed~\cite{Borghini:2001vi}, which relies on the idea that a cumulant 
expansion allows one to separate genuine collective phenomena from trivial 
short-range correlations, while they interfere in the data.
Since correlations from momentum conservation behave as short-range 
correlations, they can be removed as efficiently as them, to leave a clean 
collective flow signal.

Then, some bad news. 
Since the $M$-particle correlation from momentum conservation is of the same
order as that from short-range interactions, as for instance quantum 
correlations, it may bias measurements of these other correlations. 
It is thus worth checking that momentum conservation does not contribute 
significantly to the correlations measured by particle interferometry and 
attributed to quantum (anti)symmetrization of the wave-function. 
This is especially true when studying the dependence of HBT parameters on the 
average momentum of the particles since the correlation from momentum 
conservation increases with momentum [see Eqs.~(\ref{c2}) and (\ref{c3})].

\begin{acknowledgments}
Fruitful discussions with Fran{\c c}ois Gelis and Jean-Yves Ollitrault are 
warmly acknowledged.
\end{acknowledgments}

\end{document}